\documentclass[apj]{emulateapj}
\slugcomment{{\sc Accepted to ApJ:} June 22, 2015}

\usepackage{amssymb,amsmath}
\usepackage{natbib}
\usepackage{epstopdf}
\usepackage{graphicx}
\usepackage{floatrow}
\usepackage{rotating}
\usepackage{color}

\newcommand{\teff}{${T}_{\mathrm{eff}}$}
\newcommand{\logg}{$\log{g}$}
\newcommand{\msun}{$M_{\odot}$}

\newcommand{\Kep}{\emph{Kepler}}

\shorttitle{Outbursts and Asteroseismology of the ZZ Ceti variable KIC 4552982}
\shortauthors{Bell et al.}

\begin{document}

\title{KIC 4552982: Outbursts and Asteroseismology from the Longest \\ Pseudo-Continuous Light Curve of a ZZ Ceti}
\author{Keaton J. Bell\altaffilmark{1,2}, J. J. Hermes\altaffilmark{3}, A. Bischoff-Kim\altaffilmark{4}, Sean Moorhead\altaffilmark{1}, \\ M. H. Montgomery\altaffilmark{1,2}, Roy {\O}stensen\altaffilmark{5}, Barbara G. Castanheira\altaffilmark{1,2}, D. E. Winget\altaffilmark{1,2}}

\altaffiltext{1}{Department of Astronomy, University of Texas at Austin, Austin, TX\,-\,78712, USA}
\altaffiltext{2}{McDonald Observatory, Fort Davis, TX\,-\,79734, USA}
\altaffiltext{3}{Department of Physics, University of Warwick, Coventry\,-\,CV4~7AL, United Kingdom}
\altaffiltext{4}{Penn State Worthington Scranton, Dunmore, PA\,-\,18512, USA}
\altaffiltext{5}{Institute of Astronomy, KU Leuven, Celestijnenlaan 200D, B-3001 Heverlee, Belgium}

\email{keatonb@astro.as.utexas.edu}

\begin{abstract}
We present the \Kep{} light curve of KIC 4552982, the first ZZ Ceti (hydrogen-atmosphere pulsating white dwarf star) discovered in the \Kep{} field of view.  Our data span more than 1.5 years with a 86\% duty cycle, making it the longest pseudo-continuous light curve ever recorded for a ZZ Ceti.  This extensive data set provides the most complete coverage to-date of amplitude and frequency variations in a cool ZZ Ceti. We detect 20 independent frequencies of variability in the data that we compare with asteroseismic models to demonstrate that this star has a mass $M_* > 0.6$ \msun .  We identify a rotationally split pulsation mode and derive a probable rotation period for this star of $17.47 \pm 0.04$ hr.  In addition to pulsation signatures, the \Kep{} light curve exhibits sporadic, energetic outbursts that increase the star's relative flux by 2--17\%, last 4--25 hours, and recur on an average timescale of 2.7 days. These are the first detections of a new dynamic white dwarf phenomenon that we believe may be related to the pulsations of this relatively cool (\teff{} $=10,860\pm 120$ K) ZZ Ceti star near the red edge of the instability strip.

\end{abstract}


\section{Introduction}

As they cool, white dwarfs --- the endpoints of more than 97\% of stars in our Galaxy --- evolve through instability strips on the H-R Diagram where they pulsate due to convective driving \citep{Brickhill1991,Goldreich1999}.  ZZ Ceti variables have partially ionized hydrogen atmospheres and pulsate in the surface temperature range $12{,}600$ K $> {T}_{\mathrm{eff}} > 10{,}800$ K near the mean hydrogen-atmosphere (DA) white dwarf mass of 0.6 \msun\ \citep{Tremblay2013}. The characteristics of the pulsation modes that are excited in stars are determined by the specifics of their internal structures.  The tools of asteroseismology enable us to interpret measurements of white dwarf brightness variations to potentially constrain their masses, radii, compositions, chemical stratification, equations of state, rotation, crystallized fractions, etc.\ \citep[see reviews by][]{Winget2008,Fontaine2008,Althaus2010}.  This method allows us to conduct important investigations into the behavior of matter under the extreme physical conditions of white dwarf interiors that are beyond what is accessible for study in terrestrial laboratories.

Asteroseismology of white dwarf stars is conducted primarily through Fourier analysis of photometric light curves.  While ground-based observations of pulsating white dwarfs have been collected since 1964 \citep{Landolt1968}, the terrestrial vantage point comes with its disadvantages.  Aliasing caused by daily gaps in the data and breaks due to inclement weather can introduce artifacts to Fourier transforms (FTs).  Since spectral resolution and signal-to-noise improve with a longer observational baseline and more complete coverage, observations spanning multiple nights are often required to resolve the individual frequencies of a rich pulsation spectrum.  Efforts to surpass these limitations include extended, global observations with networks of telescopes distributed in longitude, with the greatest contributions to white dwarf science coming from the Whole Earth Telescope collaboration \citep{Nather1990}. 

More recently, the \Kep{} space mission has enabled asteroseismic research of unprecedented quality by obtaining extended time series photometry of a consistent field with a high duty cycle \citep[see, e.g.,][]{Gilliland2010a,Christensen-Dalsgaard2011}.  Besides the presently described work, detailed asteroseismic studies of pulsating white dwarfs enabled by the original \Kep{} mission include analysis of a V777 Her star \citep[helium-atmosphere variable white dwarf; KIC 8626021;][]{Ostensen2011,BischoffKim2011,Corsico2012,BischoffKim2014} and another ZZ Ceti \citep[KIC 11911480;][]{Greiss2014}.  In its present two-wheel configuration, \Kep{} has also provided asteroseismic data on the ZZ Ceti variable GD 1212 \citep{Hermes2014} and a ZZ Ceti with an M dwarf binary companion \citep{Hermes2015}. \Kep{} will observe additional white dwarf pulsators in upcoming K2 mission fields \citep{Howell2014}.

\begin{figure*}[t]
\includegraphics[width=160mm]{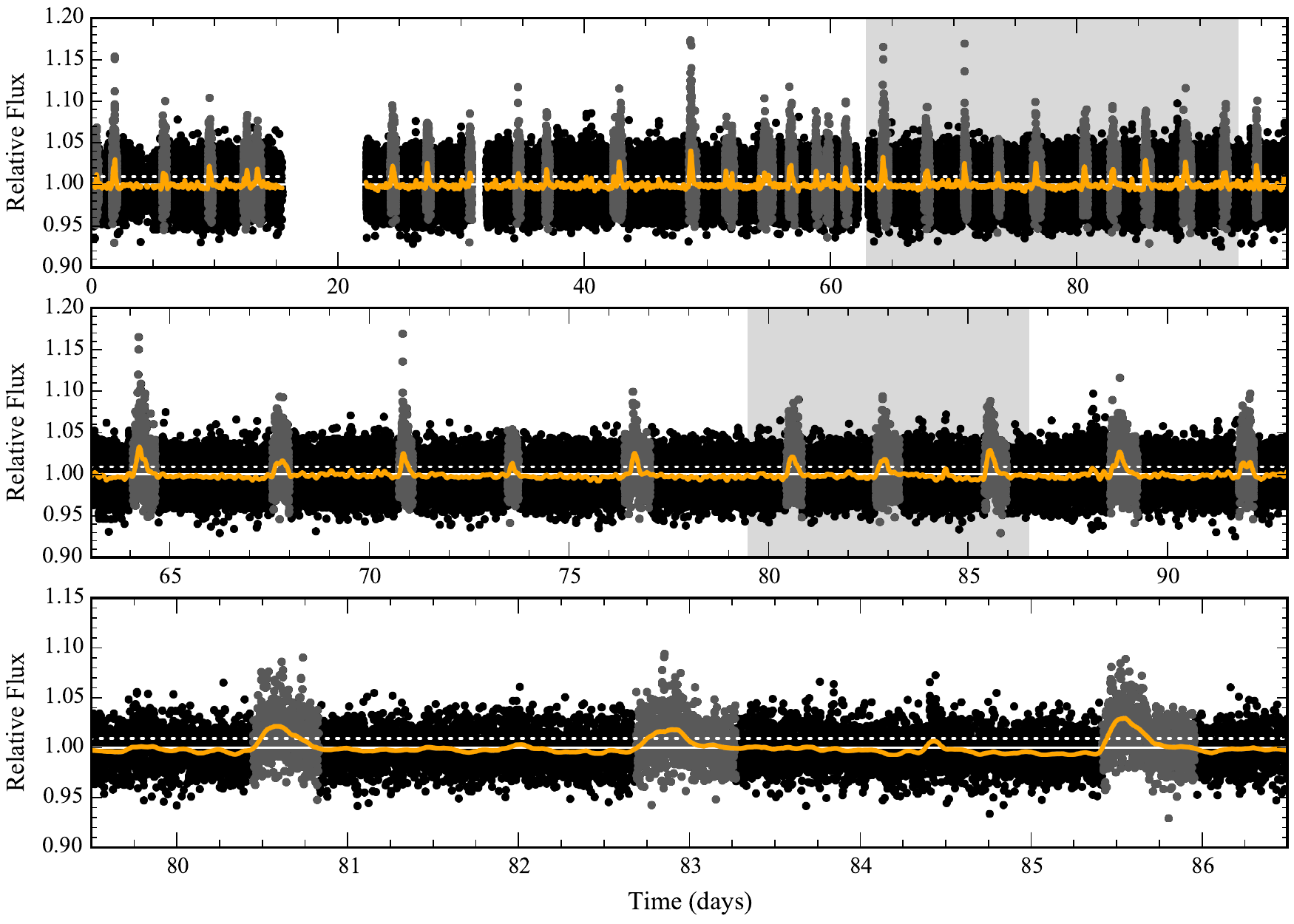}
\caption{Representative segments of the \Kep{} light curve of KIC 4552982 as a function of days since the start of Q14 observations.  The top panel shows the full Q14 light curve.  The one-month shaded region in the top panel is expanded in the middle panel.  The one-week shaded region in the middle panel is expanded in the bottom panel.  The solid line is the light curve smoothed with a 3-hour-wide Epanechnikov (inverted parabola) kernel.  The point-to-point scatter dominates the pulsation amplitudes in the light curve, so pulsations are not apparent to the eye.  The white dotted line marks the significance criterion for our outburst detection algorithm, and outbursts determined to be significant are highlighted.  We discuss this algorithm and these outburst events in Section \ref{sec:outbursts}.}
\label{fulllc}
\end{figure*}

\citet{Hermes2011} sought and discovered the first ZZ Ceti in the original \emph{Kepler} field of view:\ WD~J191643.83+393849.7.  They gathered $\sim$21 hr of time-series photometry on the 2.1-meter Otto Struve telescope at McDonald Observatory and identified seven frequencies of brightness variability, though with admittedly large uncertainties. They also obtained four low- to medium-resolution spectra for the white dwarf and fit the Balmer line profiles to models to determine its values of \teff{} $=11{,}129\pm 115$ K, \logg{} $=8.34\pm0.06$,  and $M_\star = 0.82 \pm 0.04$ \msun. Now equipped with corrective terms that take into account the effects of 3-dimensional convection \citep{Tremblay2013}, we revise the spectroscopically derived values to \teff{} $=10{,}860\pm 120$ K, \logg{} $=8.16\pm 0.06$ in this work, and interpolate the model cooling sequences of \citet{Renedo2010} to get $M_\star = 0.69 \pm 0.04$ \msun. These properties place the white dwarf at the empirical red (cool) edge of the ZZ Ceti instability strip \citep{Tremblay2013}.

This target at $19^\mathrm{h}16^\mathrm{m}43^\mathrm{s}.83$, $39^\circ 38'49.7''$ was assigned \Kep{} ID KIC 4552982 and was observed by the \Kep{} spacecraft at short cadence from Q11 until the second reaction wheel failure during Q17.  The resulting data provide the longest ($\sim$20-month) pseudo-continuous light curve of a ZZ Ceti ever obtained.  Besides resolving a rich pulsation spectrum, the \Kep{} light curve revealed a surprising outburst phenomenon unlike any previously studied white dwarf behavior.

In Section~\ref{sec:phot} we describe the methods used to optimally reduce the \Kep{} light curves.  In Section~\ref{sec:outbursts} we characterize and discuss the nature of the energetic outbursts recorded for the first time in the \Kep{} photometry.  Our asteroseismic analysis makes up Section \ref{sec:pulsations}, where we measure the pulsational properties of the star to constrain the stellar mass, hydrogen layer mass and stellar rotation rate.  We summarize our findings and conclude in Section \ref{sec:conc}.

\section{\Kep{} Photometry and Data Reduction}
\label{sec:phot}

Following the discovery by \citet{Hermes2011} that WD~J191643.83+393849.7 (KIC 4552982) pulsates, this target was prioritized by the \Kep{} Asteroseismic Science Consortium (KASC) for short-cadence monitoring by the \Kep{} spacecraft \citep[58.85 s image$^{-1}$;][]{Gilliland2010b}.  It has a recorded \Kep{} magnitude of $K_P=17.9$.  The object was observed from Quarter 11 (Q11) through Q17 (2011 September 29 to 2013 May 11).

The \Kep{} data were reduced with the {\sc PyKE} software package \citep{Still2012} following the method described by \citet{Kinemuchi2012}.  We defined custom apertures for light-curve extraction from the Target Pixel Files to maximize target signal-to-noise.  We then masked out the outburst events (described in Section 3) from the raw light curves and found that linear least-squares fits of the top six Cotrending Basis Vectors from each quarter characterized and allowed us to correct for the systematic instrumental trends in the light curve. Finally, we excluded points that were flagged for questionable data quality by the \Kep{} pipeline, and we manually removed statistical outliers falling $>5\sigma$ from the local, 10-d median, ensuring that no clipped points were related to astrophysical photometric variations. Our final light curve contains 746,916 data points and has a 86\% duty cycle. The representative Q14 light curve that we acquired after normalizing and combining each month of data is provided in Figure~\ref{fulllc}.

\section{Outbursts in the \Kep{} Light Curve}
\label{sec:outbursts}

With a mean standard deviation from measurement noise of 1.8\% dominating the pulsation amplitudes in the \Kep{} light curve, its visually striking features are the occasional large brightness enhancements that occur throughout the 20 months of observations (visible in Figure~\ref{fulllc}).  We see no evidence of a faint, red companion, and argue these observations likely mark the first detection of a new white dwarf outburst phenomenon.

\subsection{Outburst Characteristics}

\begin{figure}[b]
\includegraphics[width=\columnwidth]{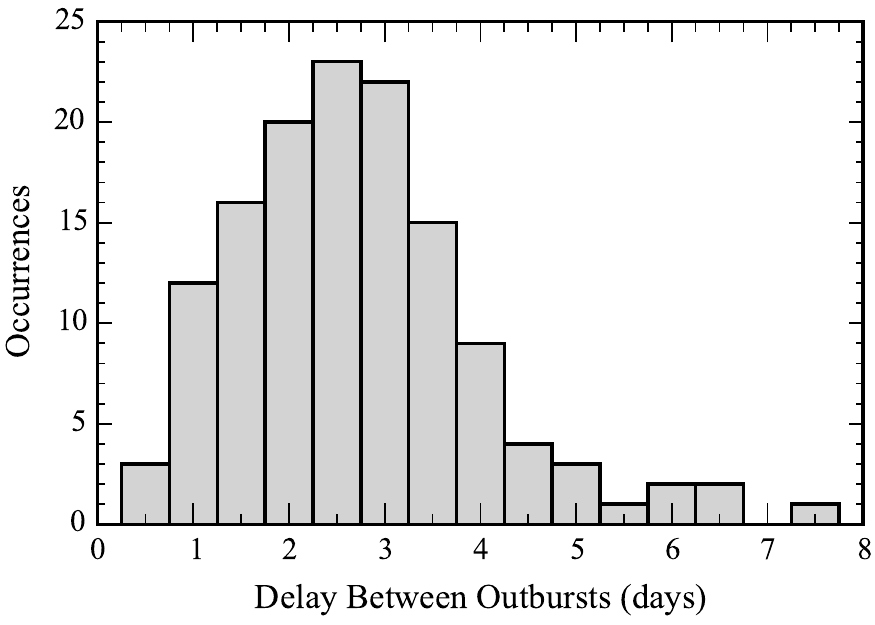}
\caption{Histogram of delay times between successive outbursts that meet our detection criteria.  The mean recurrence timescale is 2.7 days and the standard deviation is 1.3 days.}
\label{delay}
\end{figure}

We define an automatic detection algorithm to locate significant outbursts in the light curve.  Each month of data is smoothed with a 3-hour-wide Epanechnikov \citep[inverted parabola;][]{Epanechnikov1969} kernel to reduce scatter while retaining the outburst signatures.  Outbursts are identified where the light curve exceeds a significance threshold of $4 \times$ the standard deviation of the smoothed flux for 30 consecutive minutes.  We interpret the target to be in outburst until the smoothed flux returns to the median value of 1.0.  We iteratively search for outbursts by recalculating the significance threshold from ``quiescent'' (outside known outbursts) portions of the light curve until no additional outbursts are identified.  The values that we adopt for this detection scheme are tuned such that the algorithm identifies all outbursts that are obvious to the eye without selecting regions that do not stand up to human scrutiny.  Our search yields 178 events of significant (peak $2-17$\%) brightness increases that typically last $\approx 4-25$ hours.  We display the histogram of delay times between detected outbursts in Figure~\ref{delay}.  These outbursts appear to occur stochastically in time with an average delay of 2.7 days and a standard deviation of 1.3 days.

\begin{figure}[t]
\includegraphics[width=\columnwidth]{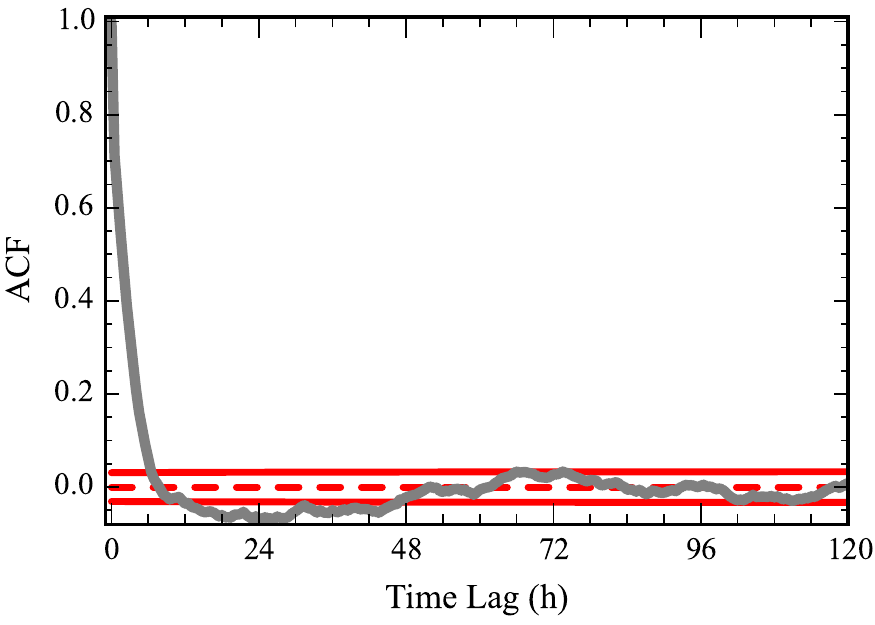}
\caption{Autocorrelation of the Q11-Q17 \Kep{} light curve of KIC 4552982.  The positive autocorrelation coefficients at short ($<8$ hour) time lag simply demonstrate that outbursts often persist to this duration.  The range of time lags that produce negative autocorrelation coefficients is interesting, as it shows that an outburst is less likely to happen within 2 days of another.  Beyond this recharge time, outbursts are consistent with being random (the solid lines are the 95\% confidence intervals for random data).}
\label{autocor}
\end{figure}

These timescale properties are reflected in the autocorrelation of the light curve, shown in Figure~\ref{autocor}.  The autocorrelation function depicts how likely it is to measure excesses in flux separated by different lags of time.  The light curve was smoothed in 30-minute bins to average over any correlation from pulsations.  The solid horizontal lines mark the 95\% confidence thresholds of $\pm 2/\sqrt{N_e}$, where $N_e$ is the effective sample size \citep{Chatfield2004}.  We approximate the effective sample size from the number of light curve bins compared at each time lag, multiplied by $(1-$ACF$_1)/(1+$ACF$_1)$, where ACF$_1$ is the autocorrelation coefficient at the smallest time lag. For a random time series, 95\% of autocorrelation coefficients would be expected to fall within these bounds.  We see positive autocorrelation at $<8$-hour time lag as this is the characteristic duration of an outburst. Beyond an apparent recharge time of roughly 2 days, outbursts are consistent with occurring at random.  The outbursts appear to be aperiodic, so we cannot formulate an event ephemeris to predict the timing of future outburst events.

\begin{figure}[t]
\includegraphics[width=\columnwidth]{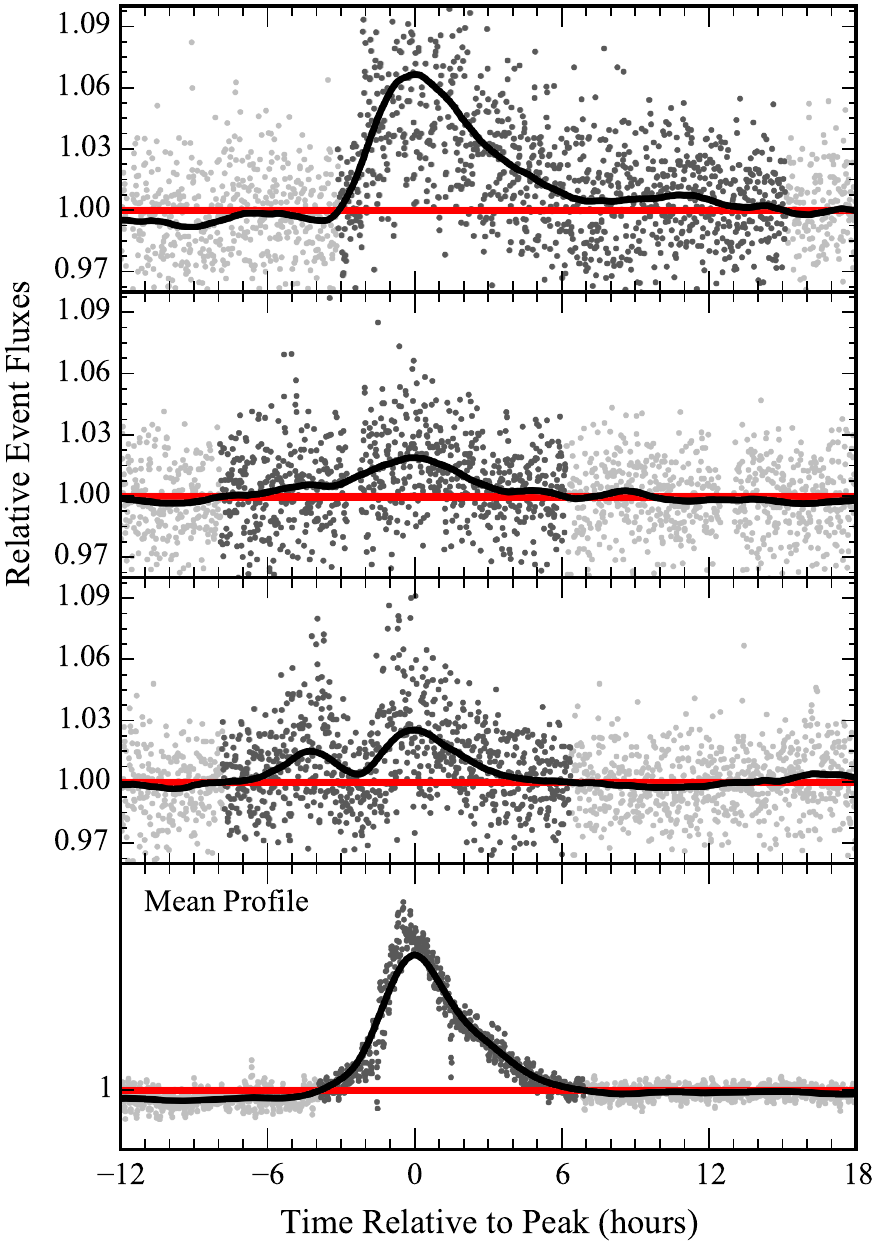}
\caption{Examples of outburst events. Top: most energetic.  Second: median energy.  Third: multi-peaked.  Bottom: mean profile.  The black curves are smoothed with a 3-hour-wide Epanechnikov kernel.  The regions determined as belonging to the outburst by our algorithm are indicated with darker gray points.}
\label{events}
\end{figure}

\begin{figure}[t]
\includegraphics[width=\columnwidth]{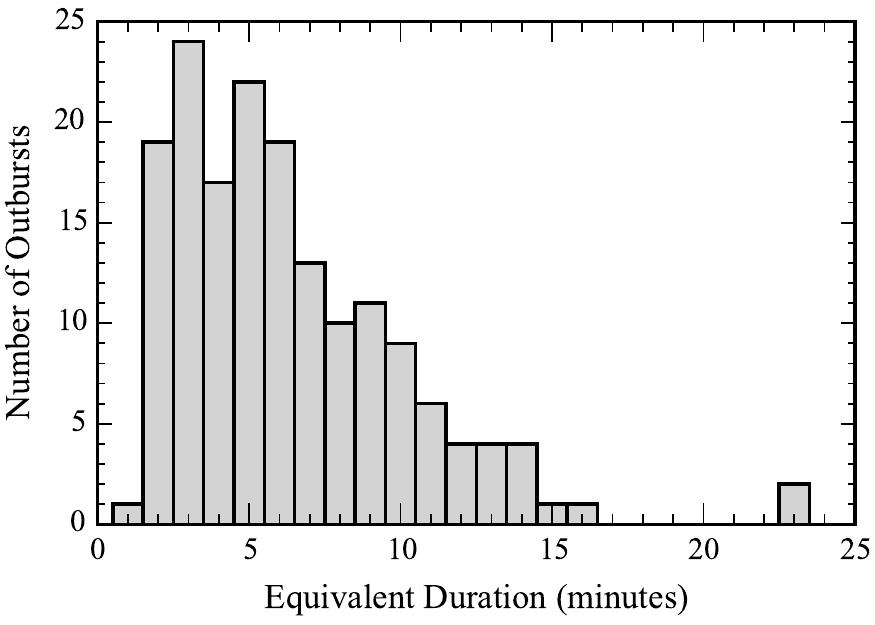}
\caption{Histogram of equivalent durations for the 167 outbursts observed in their entirety.}
\label{EDs}
\end{figure}

Figure~\ref{events}  displays regions of the light curve surrounding a few representative events in more detail.  The top panel shows the most energetic event found in the light curve, the second panel shows a median-energy event, the third panel shows a multi-peaked event, and the bottom panel displays the mean event profile that is representative of these events as a whole.  We calculated this mean profile by aligning the moments of peak brightness of all the detected events and averaging points into 58.85-s time bins.

To quantify the energies of these outbursts in the \Kep{} bandpass, we measure their equivalent durations \citep[][]{Gershberg1972}.  These are measured in the same way as spectral line equivalent widths: we integrate the flux above the local mean for the duration of the event, normalized to the mean flux level.  Assuming that the outbursts originate from the target star, this results in a value with units of time equaling the duration that the star would shine in quiescence to output the same amount of energy in the observed bandpass as the measured excess from the brightening event.  A histogram of the measured equivalent durations of the 167 outbursts that were recorded without interruption from gaps in the data is displayed in Figure~\ref{EDs}  and the continua used for the example outbursts are included in Figure~\ref{events}. The median equivalent duration of detected outbursts is 5.5 minutes (the corresponding outburst is displayed in the second panel of Figure \ref{events}). Our detection algorithm is tuned to detect large outbursts and is likely incomplete in identifying the lowest-energy events.

\subsection{Discussion of Outbursts}

The outburst characteristics that we measure are unlike any previously studied white dwarf behavior.  We argue here that these events originate from the white dwarf target KIC 4552982 and mark the first detection of a new astrophysical phenomenon.  The data in hand are insufficient for us to identify the physical mechanism that drives these events.  While we suggest possible connections to the evolutionary state of KIC 4552982, more intensive observational and theoretical explorations into their nature are left for future work. 

Each quarter of \Kep{} data was separated by rolls of the spacecraft that positioned the KIC 4552982 stellar image on a repeating sequence of four different CCD detectors.  Since the outbursts are prevalent in all quarters of \Kep{} data, they cannot be instrumental artifacts of any one of the detectors.   Furthermore, these are not widespread systematic artifacts since they do not correlate with any of the \Kep-provided Cotrending Basis Vectors that catalog trends common amongst light curves for many objects.  No other short-cadence targets show this phenomenon.  We also find no correlation between the outbursts and the short-cadence light curve that we extracted for KIC 4552992 -- a star with $K_p$ = 18.934 mag that is separated from KIC 4552982 by 12.5'' and fell mostly within the same CCD subregions read out for the target.  We note that this other source is well separated from our target and did not contribute flux to our custom photometric apertures.

From the discovery paper \citep{Hermes2011}, we have five low- to medium-resolution spectra of KIC 4552982 that cover a combined wavelength range from 4500 to 7200 \AA .  None show evidence of a companion from either radial velocity variations or spectral features that are not common of a DA (hydrogen atmosphere) white dwarf in this wavelength regime.  Since the discovery paper on KIC 4552982, new 3-dimensional convection simulations have mapped the parameters determined from 1-dimensional spectroscopic model fits to values that largely resolve a mass discrepancy in the 1-dimensional approach \citep{Tremblay2013}. We apply these corrective terms and revise the spectroscopically derived values from \citet{Hermes2011} to \teff{} $=10{,}860\pm 120$ K, \logg{} $=8.16\pm 0.06$. The white dwarf radius, as interpolated from the model cooling sequences of \citet{Renedo2010}, enables the conversion to $M_\star = 0.69 \pm 0.04$ \msun.  These parameters maintain that KIC 4552982 is one of the coolest known ZZ Ceti pulsators with a mass near or slightly above the peak of the DA white dwarf mass distribution (e.g., \citealt{Falcon2010} find $\langle M_{\rm DA}\rangle = 0.647^{+0.013}_{-0.014}$ \msun\ from gravitational redshift measurements).

\begin{figure}[t]
\includegraphics[width=\columnwidth]{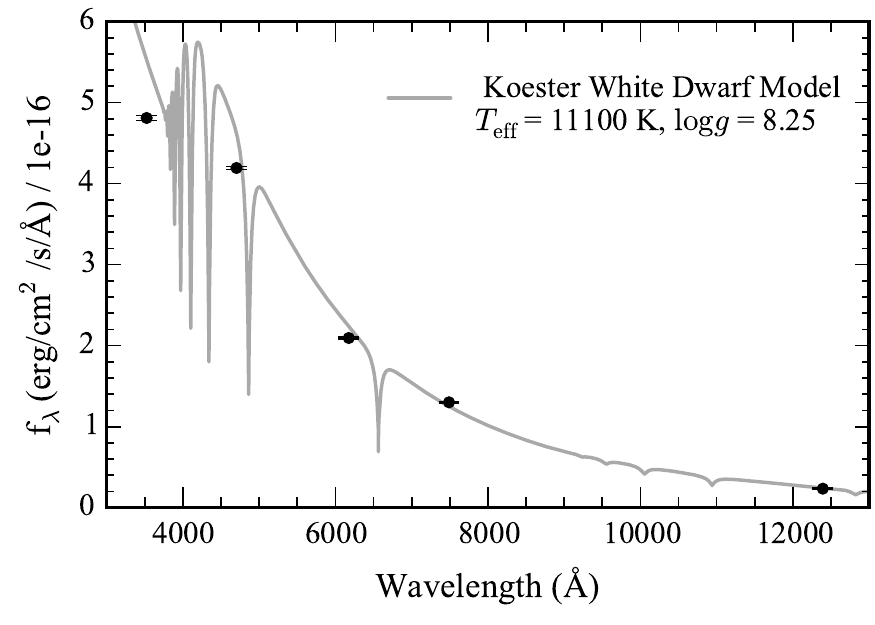}
\caption{The spectral energy distribution of KIC 4552982 in $U,g,r,i$ from the KIS survey and a $J$-band magnitude from the UKIRT J-band Public Archive.  The synthetic spectrum \citet{Koester2010} near the spectroscopically determined atmospheric parameters is plotted for reference. We see no infrared excess, allowing us to rule out a cool dwarf star companion to the white dwarf as the source of the observed outbursts.}
\label{SED}
\end{figure}

Apparent photometric magnitudes for KIC 4552928 are available from the \Kep{}-INT Survey \citep[KIS;][]{Greiss2014} in $U,g,r,$ and $i$ filters.  We include these magnitudes in Table \ref{tab:mags} along with the $J$-band magnitude acquired from the UKIRT public archive\footnote{\url{http://keplerscience.arc.nasa.gov/ToolsUKIRT.shtml}}.  All of these are reported in the Vega photometric system.  To compare these magnitudes in a spectral energy distribution (SED), we convert from the Vega system to physical AB magnitudes using the conversion factors calculated by \citet{Blanton2007} for the $U,g,r,i$ filters, and the conversion of \citet{Hewett2006} for the $J$ filter. We plot the SED of KIC 4552982 against a synthetic spectrum of a \teff{} $=11{,}000$ K, \logg{} $=8.25$ white dwarf \citep{Koester2010} for reference in Figure \ref{SED}.  We note that the photometry appears consistent with the model out to the infrared, demonstrating the absence of any cool main sequence companion that we might entertain as the possible source of the brightening events in the \Kep{} data.  We did not deredden the magnitudes, which might account for the slight discrepancy in the $U$-band.

\begin{deluxetable}{c c c}[t]
\tablecolumns{3}
\tablecaption{Apparent photometric magnitudes \\of KIC 4552982 \label{tab:mags}}
\tablehead{
\colhead{Filter} & \colhead{Vega magnitude} & \colhead{AB magnitude}}
\startdata
$U$ & $17.362 \pm 0.007$ & $18.15 \pm 0.01$ \\
$g$ & $17.755 \pm 0.005$ &  $17.68 \pm 0.01$ \\
$r$ & $17.677 \pm 0.007$ & $17.84 \pm 0.01$ \\
$i$ & $17.565 \pm 0.009$ & $17.94 \pm 0.01$ \\
$J$ & $17.76 \pm 0.03$ & $18.70 \pm 0.03$
\enddata
\end{deluxetable}

It is not likely that the outbursts are transient events from some line-of-sight flare star, rather than intrinsic to the pulsating white dwarf.  While the slightly shorter rise time than decay time of the mean outburst profile in the bottom panel of Figure~\ref{events} is qualitatively reminiscent of flares observed from low-mass main sequence stars, the observed flux increases are not nearly as rapid as the impulsive phases of classical stellar flares \citep{Moffett1974,Benz2010}.  Extensive studies of flares from early-type M dwarfs \citep[M4 and earlier;][]{Hawley2014,Davenport2014} in the short cadence \Kep{} data and from later-type M dwarfs from ground-based photometry \citep{Hilton2011} do not find such frequent, energetic flares as we detect in our data.  The durations and pulse shapes of flares from K dwarfs in the \Kep{} long-cadence data also do not match our observations \citep{Walkowicz2011}.  That the quiescent flux of a line-of-sight flare star in the \Kep{} bandpass must be orders of magnitude less than the white dwarf target to not be detected in the observed spectra or SED means that the already improbable energies/peaks/durations of the outbursts are only extreme lower-limits if they do not originate from the white dwarf.

To constrain whether the outbursts could be coming from any spatially offset faint transient source --- flare star or otherwise --- we searched for correlations between the photometric centroids within the \Kep{} images (plate scale $\approx 4''$ pixel$^{-1}$) and the occurrences of outbursts using the approach of \citet{Bryson2013}.  Our analysis was inconclusive.  Although the photometry shows significant centroid offsets when comparing the light curve during outbursts to quiescent moments, the shifts are in inconsistent directions on the sky in different quarters. When comparing just the higher-flux to lower-flux frames from quiescent segments of the light curve, an identical centroid analysis again shows significant shifts of comparable magnitude in identical directions. The measured centroid shifts of as much as $0.04''$ are likely dominated by variations in pixel sensitivity rather than the presence of any transient offset by more than of order $0.1''$.

Some white dwarfs host surface magnetic fields locally as strong as $\sim 1000$ MG \citep[e.g., PG 1031+234;][]{Latter1987}. We do not undertake any attempt to physically model the output of magnetic reconnection flares in white dwarfs for this analysis; however, the available spectra of KIC 4552982 do not show line splitting that would evidence that it is highly magnetic \citep[resolved Zeeman splitting is expected for $B \geq 10^6$ Gauss;][]{Wickramasinghe2000}.  Since this first object to exhibit this behavior is not nearly one of the more strongly magnetic white dwarfs studied, it is difficult to ascribe the outbursts to magnetic processes.  Since the timescales of dynamic events like flares should scale roughly with the dynamical timescale (free-fall time), we do not expect to see flares of multi-hour duration on a white dwarf with a $\sim 1$ s dynamical timescale.

We collected 42 hours of follow-up, higher signal-to-noise photometry with the Argos photometer \citep{Mukadam2005} on the McDonald Observatory 2.1-meter Otto Struve telescope over eight nights through a $BG40$ filter in 2013 August with hopes of catching the system in outburst. Unfortunately we made no such detection.  Given the distribution of observing windows when we were able to get useful data, there was a 47\% chance that we would have caught the peak of an outburst.  We also do not detect this phenomenon when reanalyzing the original discovery data of \citet{Hermes2011}.

\citet{DeMarco2015} analyze the \Kep{} light curve of the central star of planetary nebula Kn 61, which exhibits brightening events every 2--12 days, each lasting 1--2 days with amplitudes of 8--14\%.  These events are similar to the outbursts on KIC 4552982, at least in that they are also unlike previously studied behavior.  While the authors are unable to conclusively demonstrate a mechanism for observed outbursts in Kn 61, they explore accretion as a possible energy source.  Extending this analysis, they equate the energy of a median outburst on KIC 4552982 to $5\times 10^{-15}$ \msun\ of accreted material.  We note with some interest that this is within the range of measured asteroid masses \citep{Hilton2002}, but there is currently no known dynamical mechanism to drop circumstellar debris onto a white dwarf with this fairly regular and relatively rapid recurrence timescale \citep{Jura2008}.  Assuming a total asteroid population mass of order the Solar System asteroid belt mass \citep[$1.8 \times 10^{24}$ g;][]{Binzel2000}, this rapid accretion would deplete the mass reservoir in a mere $\sim$1000 yr.  With an outburst frequency far shorter than the calcium diffusion timescale of $\sim$40 yr for a \teff\ $=10{,}860$ K, \logg{} $=8.16$ white dwarf \citep{Koester2006}, we would expect such rapid accretion to cause strong absorption at the calcium H \& K doublet. We do not detect this signature in the available spectra \citep{Hermes2011}.

\begin{figure}[t]
\includegraphics[width=\columnwidth]{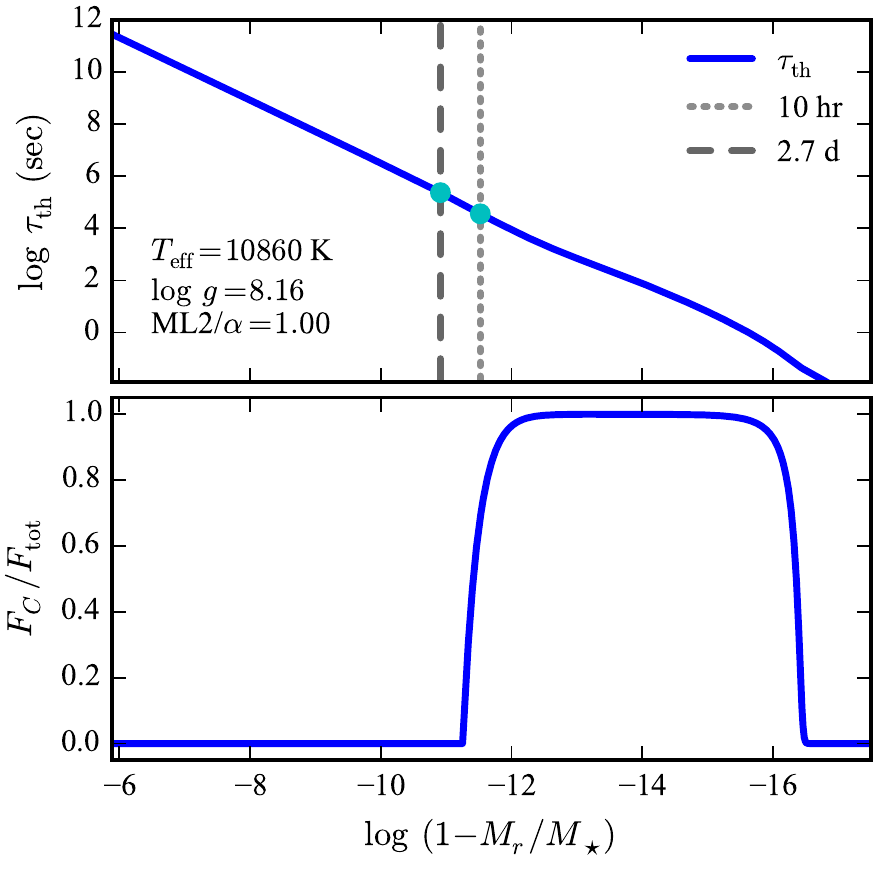}
\caption{Top: the thermal timescale as a function of depth in our model.  The 10-hr average outburst duration timescale and the 2.7-day recurrence timescale are marked with vertical dotted lines. Bottom: the fractional flux carried by convection through the static model atmosphere.  The relevant outburst timescales are of order the thermal timescale at the base of the convection zone, which is the timescale suspected to be most relevant to the cessation of pulsations at the cool edge of the instability strip.}
\label{timescale}
\end{figure}

\begin{figure*}[h!bt]
\includegraphics[width=0.9\columnwidth]{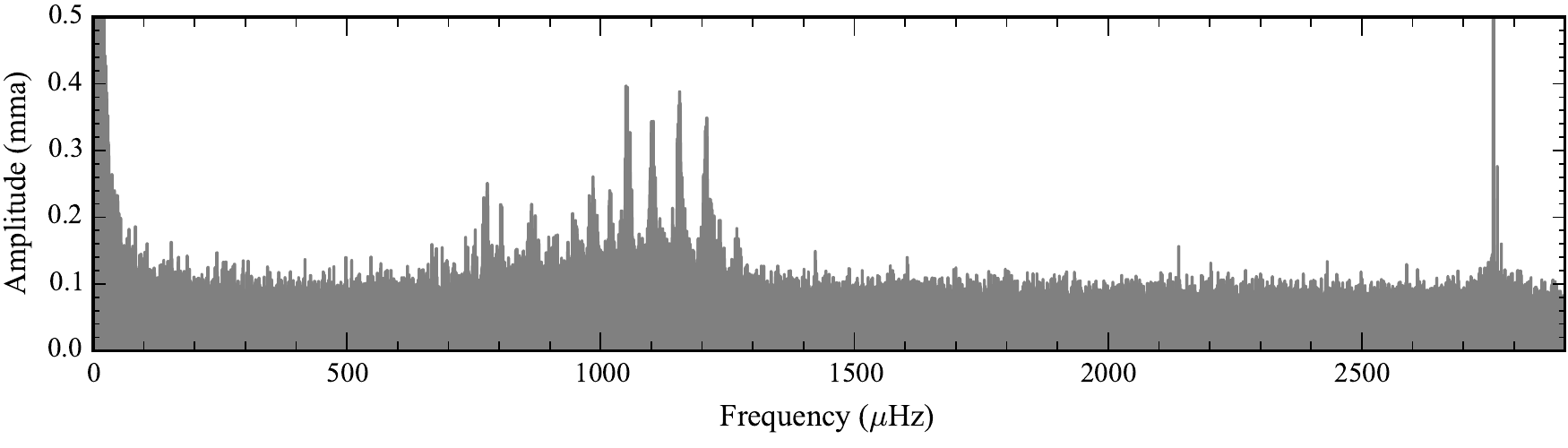}
\caption{Fourier transform of the entire Q11-Q17 \Kep{} light curve of KIC 4552982 in the region of significant pulsational variability (amplitude scale in mma = 0.1 \%).}
\label{fullft}
\end{figure*}

With \teff{} $=10{,}860\pm 120$ K, KIC 4552982 is one of the coolest ZZ Cetis known \citep{Tremblay2013}.  The mechanism that shuts down pulsations at the empirical cool edge of the ZZ Ceti instability strip is not fully understood, but may be related to the thermal timescale at the base of the convection zone exceeding some critical value \citep[e.g.,][]{VanGrootel2013}.   To explore this timescale in KIC 4552982, we use the Warsaw-New Jersey stellar envelope code \citep{Pamyatnykh1999} to calculate a static pure-hydrogen atmosphere model that matches the measured spectroscopic parameters with the ML2$/\alpha = 1.0$ mixing-length prescription for convection.  We display the run of the thermal timescale and the convection profile in the outer $10^{-6}$ of the star by mass in Figure \ref{timescale}.  We note that the thermal timescale at the base of the convection zone is of the same order as the outburst duration and recurrence timescales (and much longer than the $\sim$1 second dynamical timescale).  If this is more than just a numerical coincidence, then it is possible that this outburst behavior is a more general property of ZZ Ceti variables as they evolve out of the instability strip.

While more than $180$ ZZ Cetis have been previously discovered and studied, there has been a slight bias towards the detection of pulsators near the hot edge of the instability strip.  This is because their pulsations are exceptionally stable, enabling studies of white dwarf evolutionary cooling \citep[e.g., ][]{Kepler2005} and possible planetary companionship \citep{Mullally2008}.  By virtue of falling within the \Kep{} field, KIC 4552982 has become the most extensively observed cool-edge ZZ Ceti.  That this outburst behavior has not been previously detected in other cool-edge pulsators may be due partially to this selection bias.  The routine practice of dividing low-order polynomials from ground-based light curves to correct for changes in atmospheric extinction may also have obscured this behavior in previous observations.

\section{White Dwarf Pulsation Analysis}
\label{sec:pulsations}

The 20-month light curve yields one of the richest pulsation spectra ever resolved for a ZZ Ceti variable.  If the energetic outbursts do, in fact, originate from the white dwarf, they likely affect the stellar pulsations in a measurable way.  We explore the observed variations in pulsation mode properties that may be related to these outbursts, but unfortunately the signal-to-noise of the \Kep{} light curve is insufficient to study these variations on timescales less than the mean outburst timescale in much detail.  We proceed with an asteroseismic analysis that assumes that the outbursts do not significantly alter the locations of the peaks in the power spectrum of KIC 4552982 and reserve comments on the possible interplay between these dynamic processes for our concluding remarks.

\begin{figure*}[hbt]
\includegraphics[width=0.95\columnwidth]{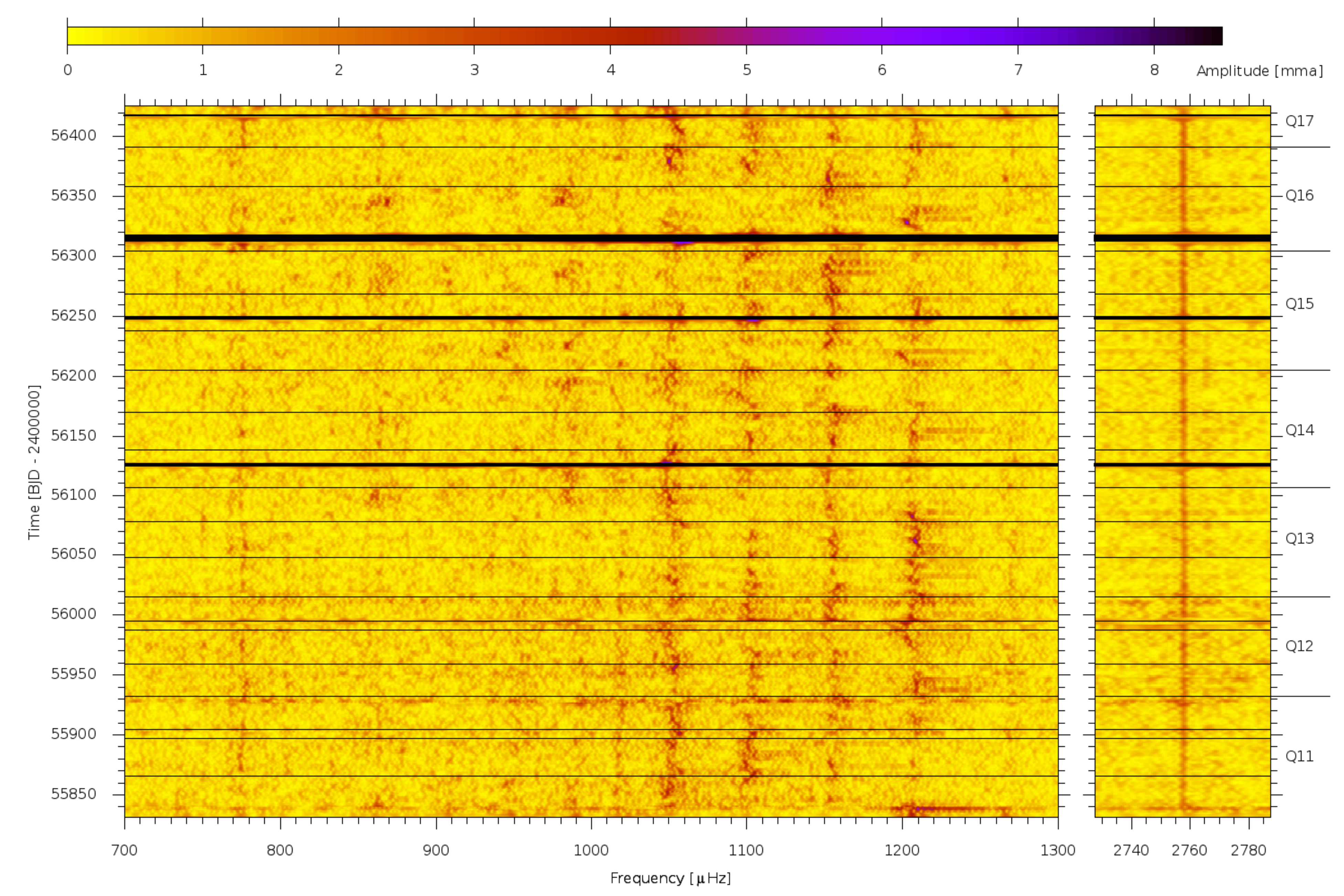}
\caption{Sliding FT of a 5-day window over the entire \Kep\ light curve.  The color bar at the top of the figure gives the amplitude scale in mma (= 0.1 \%). The modes in the 700--1300 $\mu$Hz frequency range (left panel) are observed to wander in phase (frequency) and amplitude while the mode at 2757.54 $\mu$Hz is relatively stable.}
\label{run}
\end{figure*}

\subsection{Mode Stability and Frequency Determination}
\label{sec:determination}

Figure~\ref{fullft} shows the Fourier transform (FT) of the entire Q11-Q17 light curve through the full region of significant pulsational variability.  This FT was computed with the {\sc Period04} software \citep{Lenz2004}.  We exclude all known instrumental artifacts that are harmonics of the long-cadence sampling rate of 566.41 $\mu$Hz \citep{Gilliland2010b} from this analysis.  The power at the low-frequency limit of the FT is introduced primarily by the aperiodic outbursts and mostly goes away if we remove the outburst events from the light curve before computing the FT.  Since this $1/f$ noise decays sufficiently before the low end of the frequency range of pulsations and the FT of the full light curve (including outbursts) yields an overall lower noise level in the spectrum, we do not exclude the outbursts from the light curve for our Fourier analysis.

\begin{figure*}
\includegraphics[width=0.9\columnwidth]{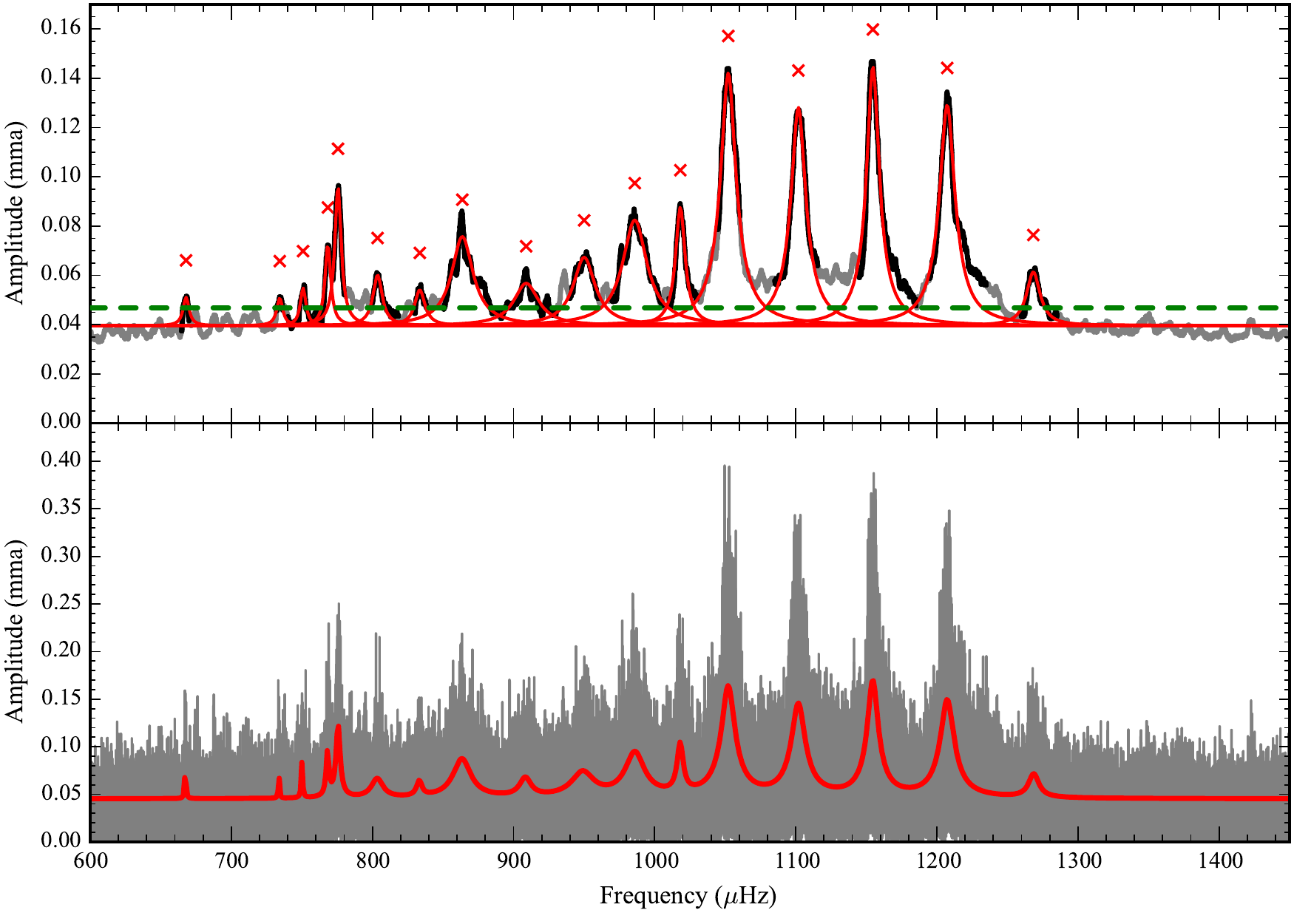}
\caption{Top: Lorentzian fits to 17 bands of pulsational power in the 4-$\mu$Hz boxcar smoothed FT.  The horizontal line at 0.0468 mma is our significance criterion.  The Lorentzians are best fits to the power (amplitude squared) above the median noise level in hand-selected (black) regions of the light curve.  
Bottom: We refine our fit parameters by simultaneously fitting these 17 bands to the original, unsmoothed FT.  The final parameters of these Lorentzians are listed in Table~\ref{tab:pers}.}
\label{powerbands}
\end{figure*}

The features of the FT of particular asteroseismic significance are the 17 bands of power in the 600--1450 $\mu$Hz frequency range and the sharp rotational triplet feature surrounding 2765.66 $\mu$Hz.  The broad power bands clearly do not reflect the sharp, 0.020-$\mu$Hz-wide spectral window (the signature of a perfect sine wave sampled identically to the data).  This is the result of modulation of the pulsation properties of the observed modes.  This modulation in amplitude and frequency is clearly seen in the sliding FT of Figure~\ref{run}.  The sliding FT is computed by sliding a 5-day window over the entire light curve and computing the FT at 1-day steps.  This time-resolved 2-dimensional FT is shown in color contour, where darker areas indicate greater power in the FT.  We observe that all the pulsations in the 700--1300 $\mu$Hz range (left panel) show rapid changes in frequency and amplitude on a timescale of a few days.  The frequency shifts of the main peaks appear to be correlated, demonstrating that these drifts cannot be ascribed purely to stochastic behavior of the individual modes.  The higher amplitudes in the sliding FT compared to the full FT in Figure~\ref{fullft} demonstrates that the instantaneous mode amplitudes are larger than the peaks of the widened power bands.  These data offer the most extensive coverage of mode variations that are typical of cool ZZ Cetis \citep[see, e.g.,][]{Pfeiffer1996, Kleinman1998, Hermes2014}.  The variations in mode properties may be related to the outbursts observed in the light curve (Section~\ref{sec:outbursts}), but we were unable to demonstrate this connection from the data. Meanwhile, the mode at 2757.54 $\mu$Hz (right panel) is relatively stable.

We interpret each band of power to be linked to a single significant pulsation mode in the star.  Since the wide nature of these bands prevents us from simply selecting and prewhitening the highest peaks in the FT, we instead adopt a method of fitting each band in the power spectrum (the squared FT) with a Lorentzian function:
\begin{equation}
\label{one}
L(\nu) = \frac{P\gamma^2}{(\nu - \nu_0)^2 + \gamma^2} + \langle FT(\nu)^2 \rangle .
\end{equation}
While Lorentzians are used to fit signatures of the stochastically driven pulsations in Sun-like stars and red giants, we emphasize that stochastic driving is unlikely to be efficient in pulsating white dwarfs since the stochastic driving timescale ($\sim$~1 second) is much shorter than the observed pulsation periods \citep[$\sim$~10 minutes;][]{Saio2013}.  We choose to fit Lorentzians as a convenience. We use slightly different methods for identifying and fitting Lorentzians to the significant frequencies in each region of the FT that are adapted to suit the different feature widths.

For the dense bands of pulsational power in the 600--1450 $\mu$Hz range, we first identify which bands are significant from the 4-$\mu$Hz boxcar-smoothed FT displayed in the top panel of Figure~\ref{powerbands}.  We use a bootstrap method to calculate a very conservative significance threshold. We randomly rearrange the light curve points 10,000 times, keeping the same time sampling.  We then calculate the FT in the 600--1450 $\mu$Hz range, apply the $4\ \mu$Hz smoothing, and record the highest value in the smoothed FT for each shuffled light curve.  We set our significance threshold equal to the 99.7 percentile in the distribution of maximum values in each shuffled, smoothed FT.  We find this value at 0.0468 mma (1 mma = 0.1\%) in the smoothed FTs.  This means there is only a $\sim 0.3\%$ chance that we would find \emph{any} peak in the smoothed FT above 0.0468 mma due to noise alone.  However, in regions near a pulsational power band, the wide feature can raise actual noise peaks above this threshold, so we conservatively fit only unambiguously significant features near larger power bands.  The peaks of the features we select are each marked with a $\times$ in the top panel of Figure~\ref{powerbands}.  Our least squares Lorentzian fits use 5 $\mu$Hz initial guesses for the HWHM at these locations (selected so that the resulting fits match the intended features by eye).  We fit Lorentzians only to the regions immediately surrounding the selected features (the black portions of the smoothed FT in the top panel of Figure~\ref{powerbands}).  The resulting Lorentzian fits to these features are displayed in the figure.

\begin{figure}[b]
\includegraphics[width=\columnwidth]{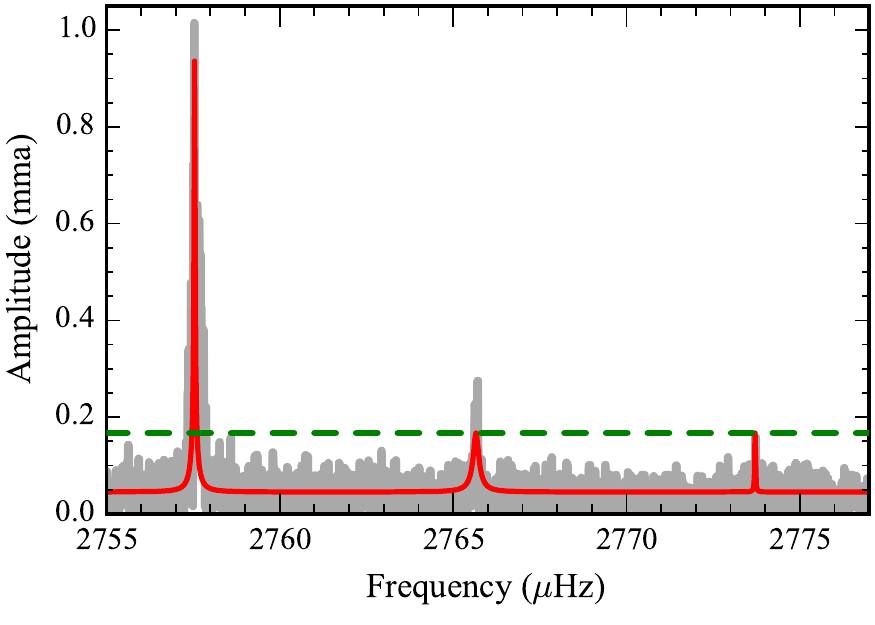}
\caption{The rotational triplet surrounding the 2765.66 $\mu$Hz mode. The power was fit with the displayed set of Lorentzians.  The highest-frequency component consists of a single point above the noise that is fit to its peak power by the Lorentzian.}
\label{triplet}
\end{figure}

While the smoothed FT is useful for selecting significant peaks in frequency space, the smoothing can affect the HWHMs of the Lorentzian fits.  Since these HWHMs may have an astrophysical significance, we refine our fit parameters by fitting to the unsmoothed FT.  We refit all of the Lorentzians concurrently with a common bias-level parameter to fit the average noise level.  The result of this fit is plotted over the unsmoothed FT in the bottom panel of Figure~\ref{powerbands}.  Because these power bands encompass a forest of high- and low- amplitude peaks, the final Lorentzians fit through the average amplitudes and do not reach the maximum peaks of the features. The central frequencies, HWHMs, and heights of the Lorentzian fits are listed with the central mode periods in Table~\ref{tab:pers}. Their relation to Equation~\ref{one} is also indicated.

\begin{deluxetable}{l c c c c}
\tablecolumns{5}
\tablecaption{Properties of Lorentzian fits to significant frequencies of pulsational variability in the FT. \label{tab:pers}}
\tablehead{
\colhead{Mode} & \colhead{Period} & \colhead{Frequency} & \colhead{HWHM} & \colhead{Lorentzian height} \\ & \colhead{(s)} & \colhead{($\nu_0$; $\mu$Hz)} & \colhead{($\gamma$; $\mu$Hz)} & \colhead{($P$; mma$^2$)} }
\startdata
$f_{1}\tablenotemark{a}$ & 1498.32 & 667.42 & 0.34 & 0.0062\\
$f_{2}$ & 1362.95 & 733.70 & 0.26 & 0.0056\\
$f_{3}\tablenotemark{a}$ & 1333.18 & 750.09 & 0.46 & 0.0050\\
$f_{4}\tablenotemark{ab}$ & 1301.73 & 768.21 & 0.97 & 0.0070\\
$f_{5}\tablenotemark{ab}$ & 1289.21 & 775.67 & 1.38 & 0.0132\\
$f_{6}\tablenotemark{a}$ & 1244.73 & 803.38 & 4.50 & 0.0023\\
$f_{7}\tablenotemark{a}$ & 1200.18 & 833.21 & 2.15 & 0.0018\\
$f_{8}\tablenotemark{a}$ & 1158.20 & 863.41 & 6.59 & 0.0055\\
$f_{9}$ & 1100.87 & 908.38 & 4.51 & 0.0023\\
$f_{10}$ & 1053.68 & 949.06 & 8.61 & 0.0032\\
$f_{11}$ & 1014.24 & 985.96 & 7.13 & 0.0066\\
$f_{12}$ & 982.23 & 1018.09 & 2.21 & 0.0081\\
$f_{13}\tablenotemark{a}$ & 950.45 & 1052.13 & 4.58 & 0.0246\\
$f_{14}\tablenotemark{a}$ & 907.59 & 1101.82 & 4.75 & 0.0189\\
$f_{15}\tablenotemark{a}$ & 866.11 & 1154.59 & 3.68 & 0.0266\\
$f_{16}\tablenotemark{a}$ & 828.29 & 1207.31 & 4.67 & 0.0202\\
$f_{17}\tablenotemark{a}$ & 788.24 & 1268.65 & 3.93 & 0.0029\\
$f_{18}\tablenotemark{ab}$ & 362.64 & 2757.54 & 0.013 & 0.8743\\
$f_{19}\tablenotemark{ab}$ & 361.58 & 2765.66 & 0.007 & 0.0260\\
$f_{20}\tablenotemark{ab}$ & 360.53 & 2773.71 & 0.009 & 0.0260\tablenotemark{c}
\enddata
\tablenotetext{a}{Likely part of the $\ell=1$ sequence based on alignment with the mean $\ell=1$ period spacing or observed rotational splitting.}
\tablenotetext{b}{Likely rotationally split components of an $\ell=1$ mode.}
\tablenotetext{c}{This fit matches the single peak for this component in the power spectrum with a height corresponding to peak power.}
\end{deluxetable}

Figure \ref{triplet} shows a higher-frequency region of the original, unsmoothed FT.  Here we observe a triplet of significant variability with near-even frequency spacing.  We calculate a significance threshold of 99.7\% confidence at 0.167 mma by a bootstrap approach similar to our calculation for the lower-frequency power bands.  The difference between the 0.167 mma significance threshold here and the 0.0468 mma threshold for the smoothed FT at lower frequency results purely from the boxcar smoothing of the latter.  While only the two lowest-frequency components rise above this significance threshold, we relax our criterion for the third peak because it falls in step with the others as is expected for rotational splitting (see Section \ref{sec:rotation}) and is only barely below our very conservative signifiance threshold.  We simultaneously fit the three features with Lorentzians as we did for the wider power bands using 1 $\mu$Hz for the initial guesses for the HWHM and adopting the bias level fit in the previous region.  The third component consists of a single peak above the noise, and the height of the Lorentzian fits the peak power of that mode. These best-fit parameters are included in Table \ref{tab:pers}.

We adopt these 20 frequencies as the mean pulsation mode frequencies of KIC 4552982.  Since we used an unconventional method for determining these mode parameters for pulsating white dwarfs,  it is difficult to assign absolute uncertainties.  We advise that researchers wishing to fit their own asteroseismic models to these frequencies weight their fits by $1/$HWHM$^2$ for each mode. Since we resolve the frequency variations over our extended light curve, our frequency determinations should be far more accurate than the HWHM.

\subsection{Period Spacing}
\label{sec:spacing}

The periods of the nonradial $g$-mode pulsations characteristic of white dwarfs of a given degree ($\ell$) and sequential radial order ($k$) reach an even spacing at the asymptotic limit of high $k$. Because of geometric cancellation, we expect the pulsations we detect to be of degree $\ell =1$ or $\ell = 2$ \citep{Dziembowski1977}.  With such a rich pulsation spectrum, we can hope to sample these sequences sufficiently well to determine the mean period spacing with radial order of one or both of these sequences.

\begin{figure}[t]
\includegraphics[width=\columnwidth]{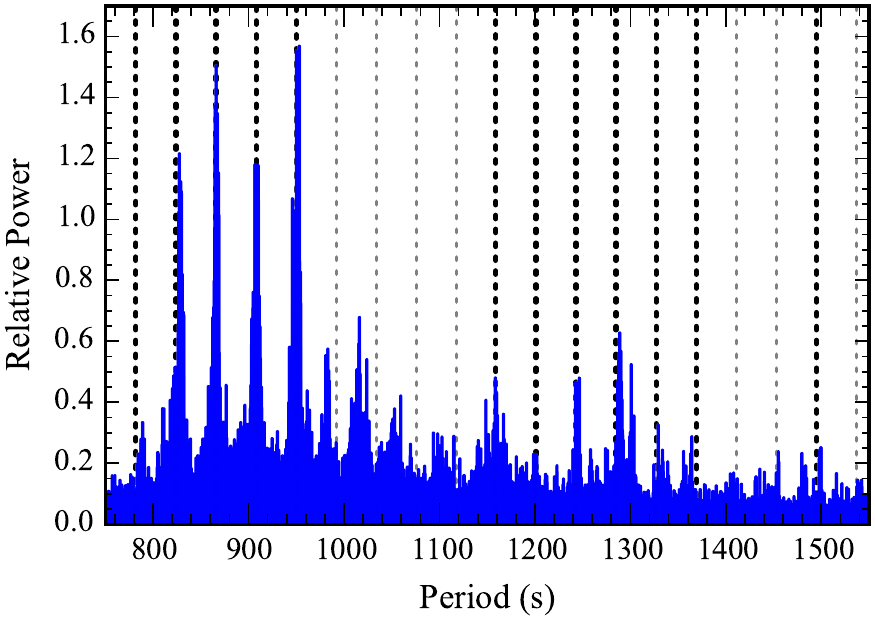}
\caption{The period transform of the full \Kep{} light curve in the region of observed pulsational power bands.  The locations of the expected $\ell = 1$ modes from the mean period spacing determined in Figure \ref{periodspacing} are marked as dotted vertical lines.  These lines are drawn darker where they fall within 7 seconds of one of our measured mode periods listed in Table \ref{tab:pers}}
\label{periodtransform}
\end{figure}

\begin{figure}[t]
\includegraphics[width=\columnwidth]{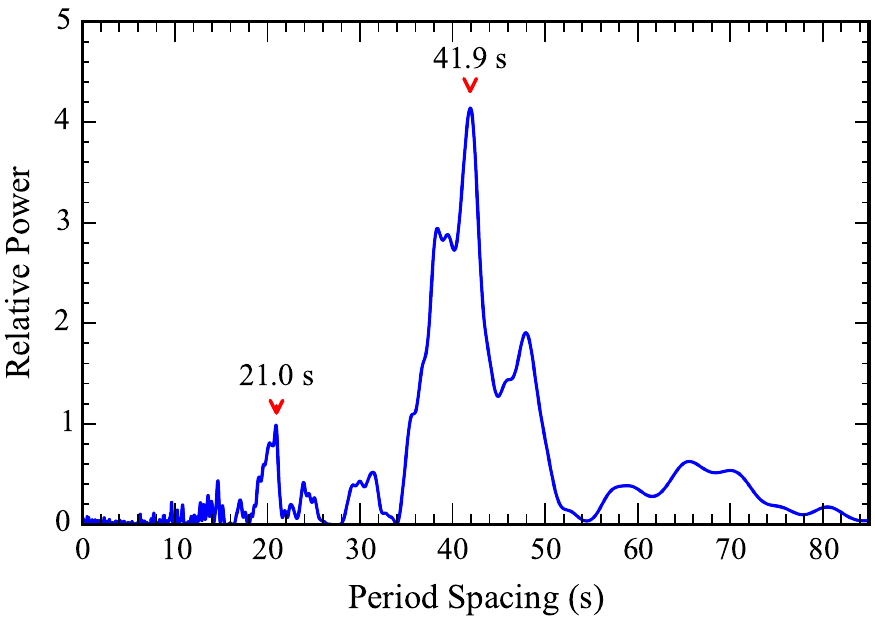}
\caption{The period transform of the period transform reveals a significant mean period spacing at 41.9 seconds and its harmonic.  This likely corresponds to a mean period spacing of $\ell =1$ modes.}
\label{periodspacing}
\end{figure}

We approach this measurement by taking the period transform of the period transform of the entire light curve as is described in its application to the DOV (pulsating hot pre-white dwarf) star PG 1159-035 in \citet[][Section 5.2.1]{Winget1991}. We emphasize that this approach is independent of the period determination in Section~\ref{sec:determination}. We arrive at the period transform by simply inverting the x-axis of the FT in the region of dense pulsational power bands (between 500 and 1500 $\mu$Hz, corresponding to a period range 667--2000 s).  This intermediate result is shown in Figure \ref{periodtransform}.  If the signals from pulsations correspond to one or more sequences with evenly spaced periods, the period transform of this period transform should reveal the spacings.  We show the resulting power spectrum in Figure \ref{periodspacing} and mark two peaks of interest: a significant spacing at $41.9\pm0.2$ s and its overtone at $20.97\pm 0.02$ s (uncertainties determined from 100 Monte Carlo simulations with {\sc Period04}).  This overtone arises from the decidedly non-sinusoidal nature of the periodic signal in the period transform.  Using the phase information we get from least-squares sinusoidal fits, we mark the locations of expected modes in the even period-spacing sequence as vertical dotted lines in Figure \ref{periodtransform} and indicate those modes that fall within 7 seconds of the measured spacing with darker dotted lines and with footnotes in Table~\ref{tab:pers}.

\subsection{Comparison with Asteroseismic Models}

We consider a grid of more than $14{,}000$ DA white dwarf cooling models that we calculated from the White Dwarf Evolution Code \citep[WDEC;][]{Lamb1975, Wood1990} following the treatment described in \citet{BischoffKim2008}.  We vary the following three parameters that most influence the mean period spacing, with noted resolution:  $10{,}000$ K $\le$ \teff\ $\le 12{,}000$ K ($200$ K resolution), 0.500 \msun\ $\le M_* \le 0.800 $ \msun\  (0.005 \msun\ resolution), and $-6.00 \le \log{M_{\rm H}/M_\star} \le -4.00$ (in steps of 0.10). This set of models widely encompasses the spectroscopic $M_*$, \teff\ values of KIC 4552982. Each of these models has a helium layer mass of $\log{M_{\rm He}/M_\star}= -2.00$ \citep[close to the value calculated for 0.7 \msun\ white dwarfs by][]{Lawlor2006}.  The core profiles have central abundances of 30\% carbon and 70\% oxygen and are homogenous out to 0.5 $M_r/M_*$ \citep[chosen to be in good agreement with][]{Salaris1997}.

\begin{figure}[t]
\includegraphics[width=\columnwidth]{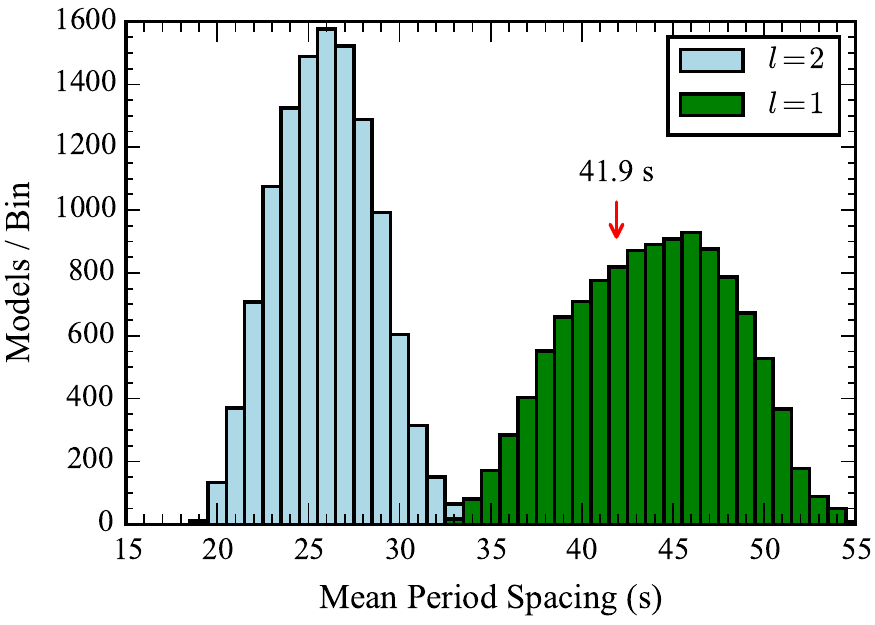}
\caption{Histograms of mean period spacings calculated for $\ell=1$ and $\ell=2$ sequences throughout our model grid.  The measured spacing for KIC 4552982 of 41.9 seconds fits exclusively within the $\ell=1$ distribution. \\ \\}
\label{periodspacinghist}
\end{figure}

\begin{figure}[t]
\includegraphics[width=\columnwidth]{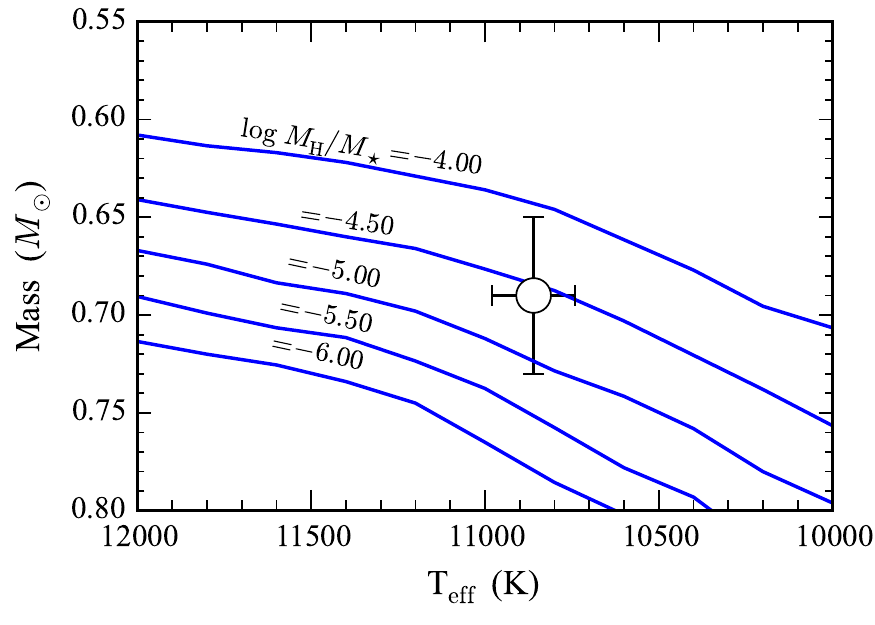}
\caption{The locations of models through $M_*$ -- \teff\ space with mean $\ell=1$ period spacings equal to the measured spacing of 41.9 seconds. The individual runs of solutions correspond to $\log{M_{\rm H}/M_\star}$ values from -4.00 (bottom track) to -6.00 (highest track) in 0.50 resolution. All models that fit the observations have $M_* > 0.60$ \msun. The spectroscopically determined parameters are overplotted in white, and are in slightly better agreement with models nearer the high-$M_{\rm H}$ end.}
\label{hmass}
\end{figure}

The measured mean period spacing of $41.9\pm0.2$ s fits well within the distribution of calculated mean spacings in our models for $\ell=1$ modes.  For $\ell=2$ modes, the models predict mean spacings nearer to 26 s.  These distributions are compared in Figure \ref{periodspacinghist} . 

We suggest that the modes that fall in step with the measured mean period spacing are likely part of the $\ell=1$ sequence with footnotes in Table~\ref{tab:pers}. Undoubtedly, some $\ell=2$ modes may by chance fall in line with this sequence and be misidentified.  Mode trapping and the dominance of different rotationally split components of the modes (see Section~\ref{sec:rotation}) can shift individual frequencies significantly from an equal period spacing, so some of the modes that we do not observe to closely match the mean spacing may also have $\ell=1$.  The dominance of different rotationally split components should not greatly affect our asteroseismic inferences \citep{Metcalfe2003}.

We display the run of models with mean $\ell=1$ spacings of 41.9 s through $M_*$, \teff\ space for selected hydrogen layer masses in our model grid in Figure \ref{hmass}.  The mean period spacing of $g$-modes is most sensitive to the overall mass, and our model comparison demonstrates asteroseimically that KIC 4552982 is likely more massive than 0.6 \msun --- in good agreement with the spectroscopic mass determination.

The spectroscopic point is plotted with error bars over the $\log{M_{\rm H}/M_{\star}}$ tracks for reference.  We observe that while the uncertainty in the spectroscopically determined $M_*$ spans multiple tracks at the resolution of our model grid, the spectroscopic point agrees best with a $\log{M_{\rm H}/M_\star}$ value of -4.70 and generally falls nearer to the maximal $\log{M_{\rm H}/M_\star}$ extent of our model grid at -4.00.

\subsection{Rotational Splitting}
\label{sec:rotation}

At higher frequency than the region of dense pulsational power bands, we noted three modes split evenly in frequency (Figure \ref{triplet}).  Even frequency spacing can result from rotational splitting of a mode into $2\ell+1$ components \citep{Unno1989}.   While we emphasize that our absolute uncertainties in the individual frequencies determined in Section \ref{sec:determination} are smaller than the HWHM of our Lorentzian fits, we will treat the HWHM here as proxies for uncertainty.  From these values, we calculate a mean frequency splitting of $8.09 \pm 0.02\ \mu$Hz.

With only three components in the rotational multiplet, we are unable to definitively assign a spherical degree, $\ell$, to this mode, so we consider the three most likely possibilities: these are either all $m=-1,0,1$ components of a $\ell=1$ triplet, three consecutive components of a $\ell=2$ quintuplet, or the $m=-2,0,2$ components of a $\ell=2$ quintuplet.

We derive first-order analytical rotation rates under the assumption that the rotation period is much longer than the period of the split mode and that the white dwarf rotates as a solid body.  We use the relation that $\Delta\nu_{k\ell m} = (m/P_{rot})\times (1-C_{k\ell})$ \citep{Brickhill1975} for solid body rotation.  At the limit of high $k$ ($k \ga 10$), $C_{k\ell}$ approaches $1/\ell (\ell+1)$.  If we assume that this is an $\ell = 2$ mode, we are safely in the asymptotic limit ($k \sim\ 11$ or 12 according to our models).  If these are adjacent splittings of a $\ell=2$ mode (i.e., $m=-2,-1,0$, $m=-1,0,1$ or $m=0,1,2$), we derive a rotation period of $28.63 \pm 0.07$ hours.   If these are $\ell=2$, $m=-2,0,2$ modes, we get twice that rotation period at $P_{rot} = 57.26 \pm 0.13$ hours.  If this is the rotationally split triplet of a $\ell=1$ mode, we cannot safely assume that these modes are in the asymptotic limit.  We guide our interpretation for this case by comparing the mode periods to our asteroseismic model with parameters \teff\ $= 10{,}800$ K and $M_* = 0.700 $ \msun\, and $\log{M_{\rm H}/M_\star} = -4.70$.  This model most closely matches our measured mean $\ell=1$ period spacing and our spectroscopically determined atmospheric parameters. The measured 361.58-s period measured for the central component of the triplet falls between the following two $\ell=1$ modes in the model: $k=5$ at 311.6 s with $C_{k\ell} =  0.491$, and $k=6$ at 379.0 s with $C_{k\ell} =  0.492$.  These $C_{k\ell}$ values essentially match the asymptotic result, and we adopt the mean of these values to calculate a rotational period of $17.47 \pm 0.04$ hours.  These periods are all plausible considering the range of asteroseismically determined rotation periods of other white dwarfs \citep{Kawaler2004}, and they support our assumption that the rotation period is much longer than the pulsation periods.

Revisiting the region of wide pulsational power bands between 600--1450 $\mu$Hz (Figure \ref{powerbands}), we see suggestions of structure at a similar $\approx 8\ \mu$Hz spacing.  This includes a pair of frequencies that passed our significance criteria and are indicated with a footnote in Table \ref{tab:pers}: $f_4, f_5$.  Since these bands fall in line with the $\ell=1$ mean period spacing pattern from our analysis in Section \ref{sec:spacing}, we prefer the $\ell=1$ rotation period of $17.47 \pm 0.04$ hours for this star.

\section{Conclusions and Future Work}
\label{sec:conc}

The \Kep{} light curve for KIC 4552982 exhibits two features of great interest: clear outburst phenomena and a rich spectrum of pulsations with frequency and amplitude modulations.  We argue that these outbursts likely originate from the pulsating white dwarf and are the first observations of a new astrophysical phenomenon.  Such energetic events likely affect the pulsations in a measurable way.  Owing to the low signal-to-noise of the \Kep{} photometry of this faint target, we are unable to study changes in the pulsation spectrum strictly before and after the observed outburst events.

The most compelling evidence for the outbursts affecting the pulsations is the relative sharpness of the triplet of high-frequency modes surrounding $2765.66\ \mu$Hz compared to the wide bands of power at lower frequency.  If these modes are all of the same degree, $\ell$, the lower-frequency, higher-$k$ modes are more sensitive to regions nearer the surface of the star.  We demonstrate that most of the modes we observe are likely part of a $\ell = 1$ sequence. The detected relative unsteadiness of the lower-frequency modes may then be an indication that the outbursts are a surface phenomenon \citep[a similar argument is made for solar-type pulsators by][]{Karoff2014}. However, variations in the pulsation frequencies and amplitudes are commonly observed in cool ZZ Cetis, so this is not the only viable interpretation.  If outbursts of the type we see in KIC 4552982 turn out to be common of cool ZZ Cetis, they may be related to the frequency modulations observed in these other stars. 

We updated the spectroscopic parameters of \citet{Hermes2011} with corrections from 3D convective simulations to get \teff{} $=10{,}860\pm 120$ K, \logg{} $=8.16\pm 0.06$, and $M_\star = 0.69 \pm 0.04$ \msun.  These parameters place KIC 4552982 at the extreme cool edge of the ZZ Ceti instability strip where pulsations are just shutting down.  We demonstrate that the average duration ($\approx 10$ hours) and recurrence ($\approx 2.7$ days) timescales observed for the outbursts are of order the thermal timescale at the base of the convection zone and suggest that these outbursts could be a feature common to all white dwarfs with this convection zone depth, and therefore temperature.

We identify 20 independent pulsation frequencies including rotationally split components as characterized in Table \ref{tab:pers}.  Comparing the measured mean period spacing of the nearly sequential $k$-overtones of likely $l=1$ modes to our models, the asteroseismology supports the spectroscopic finding for the white dwarf mass: $M_* > 0.6$ \msun.  These spectroscopic parameters match our asteroseismic models with $41.9\pm0.2$ s mean period spacings slightly better closer to the thick outer hydrogen layer end of our grid ($\log{M_{\rm H}/M_\star} = -4.00$) rather than the thin limit of our models ($\log{M_{\rm H}/M_\star} = -6.00$). We also derive a likely rotation period of $17.47 \pm 0.04$ hours if the observed modes surrounding 2765.66 $\mu$Hz are a rotationally split $\ell=1$ triplet.

The sheer extent of this nearly continuous 20-month light curve has provided an unparalleled look at dynamic white dwarf processes.  Continued asteroseismic interpretation of the identified pulsation frequencies and outburst characteristics will hopefully lead to additional constraints on the internal structure of KIC 4552982.

\acknowledgements
We thank the anonymous referee for insightful feedback that helped to improve the impact of the paper. K.J.B. thanks Michel Breger and Patrick Lenz for fruitful discussions and custom modifications of the Period04
Fourier analysis software \citep{Lenz2004}; Detlev Koester for providing a white dwarf model spectrum; Sandra Greiss for providing \Kep{}-INT data; Patrick Kelly for observing assistance; Dave Doss, Coyne Gibson and John Kuehne for observing support; and Jim Davenport, Suzanne Hawley and Eric Hilton for their insights into M dwarf flare frequency distributions in \Kep{}.  We also thank KASC Working Group 11 for submitting this target as a highly ranked short-cadence target, making possible the extensive light curve analyzed here.  This research was funded in part by the NSF grants AST-0909107 and AST-1312983, grant 003658-0252-2009 from the Norman Hackerman Advanced Research Program, the \Kep{} Cycle 4 GO proposal 11-KEPLER11-0050, and NASA grant NNX12AC96G.  J.J.H. additionally acknowledges funding from the European Research Council under the European Union's Seventh Framework Programme (FP/2007-2013) / ERC Grant Agreement n. 320964 (WDTracer). This paper includes data collected by the Kepler mission. Funding for the Kepler mission is provided by the NASA Science Mission Directorate. Some of the data presented in this paper were obtained from the Multimission Archive at the Space Telescope Science Institute (MAST). STScI is operated by the Association of Universities for Research in Astronomy, Inc., under NASA contract NAS5-26555. Support for MAST for non-HST data is provided by the NASA Office of Space Science via grant NAG5-7584 and by other grants and contracts. This work made use of PyKE \citep{Still2012}, a software package for the reduction and analysis of Kepler data. This open source software project is developed and distributed by the NASA Kepler Guest Observer Office. The United Kingdom Infrared Telescope (UKIRT) is supported by NASA and operated under an agreement among the University of Hawaii, the University of Arizona, and Lockheed Martin Advanced Technology Center; operations are enabled through the cooperation of the Joint Astronomy Centre of the Science and Technology Facilities Council of the U.K.  When the some of the data reported here were acquired, UKIRT was operated by the Joint Astronomy Centre on behalf of the Science and Technology Facilities Council of the U.K. The authors acknowledge the Texas Advanced Computing Center (TACC) at The University of Texas at Austin for providing data archiving resources that have contributed to the research results reported within this paper.


\begin{thebibliography}{}
\bibitem[Althaus et al.(2010)]{Althaus2010} Althaus, L.~G., C{\'o}rsico, A.~H., Isern, J., \& Garc{\'{\i}}a-Berro, E.\ 2010, \aapr, 18, 471 
\bibitem[Benz \& G{\"u}del(2010)]{Benz2010} Benz, A.~O., \& G{\"u}del, M.\ 2010, \araa, 48, 241 
\bibitem[Binzel et al.(2000)]{Binzel2000} Binzel, R.~P., Hanner, 
M.~S., \& Steel, D.~I.\ 2000, Allen's Astrophysical Quantities, 315 
\bibitem[Bischoff-Kim et al.(2008)]{BischoffKim2008} Bischoff-Kim, A., Montgomery, M.~H., \& Winget, D.~E.\ 2008, \apj, 675, 1505 
\bibitem[Bischoff-Kim \& {\O}stensen(2011)]{BischoffKim2011} Bischoff-Kim, A., \& {\O}stensen, R.~H.\ 2011, \apjl, 742, L16 
\bibitem[Bischoff-Kim et al.(2014)]{BischoffKim2014} Bischoff-Kim, A., {\O}stensen, R.~H., Hermes, J.~J., \& Provencal, J.~L.\ 2014, \apj, 794, 39 
\bibitem[Blanton \& Roweis(2007)]{Blanton2007} Blanton, M.~R., \& Roweis, S.\ 2007, \aj, 133, 734 
\bibitem[Brickhill(1975)]{Brickhill1975} Brickhill, A.~J.\ 1975, \mnras, 170, 405 
\bibitem[Brickhill(1991)]{Brickhill1991} Brickhill, A.~J.\ 1991, \mnras, 251, 673 
\bibitem[Bryson et al.(2013)]{Bryson2013} Bryson, S.~T., Jenkins, J.~M., Gilliland, R.~L., et al.\ 2013, \pasp, 125, 889 
\bibitem[Chatfield(2004)]{Chatfield2004} Chatfield, C.\ 2007, The Analysis of Time Series: An Introduction, (6th ed.; Florida, US: CRC Press)
\bibitem[Christensen-Dalsgaard \& Thompson(2011)]{Christensen-Dalsgaard2011} Christensen-Dalsgaard, J., \& Thompson, M.~J.\ 2011, IAU Symposium, 271, 32 
\bibitem[C{\'o}rsico et al.(2012)]{Corsico2012} C{\'o}rsico, A.~H., Althaus, L.~G., Miller Bertolami, M.~M., \& Bischoff-Kim, A.\ 2012, \aap, 541, A42 
\bibitem[Davenport et al.(2014)]{Davenport2014} Davenport, J.~R.~A., Hawley, S.~L., Hebb, L., et al.\ 2014, \apj, 797, 122 
\bibitem[De Marco et al.(2015)]{DeMarco2015} De Marco, O., Long, J., Jacoby, G.~H., et al.\ 2015, \mnras, 448, 3587 
\bibitem[Dziembowski(1977)]{Dziembowski1977} Dziembowski, W.\ 1977, Acta Astron., 27, 1 
\bibitem[Epanechnikov(1969)]{Epanechnikov1969} Epanechnikov, V.~A.\ 1969, Theory Probab. Appl., 14, 153
\bibitem[Falcon et al.(2010)]{Falcon2010} Falcon, R.~E., Winget, D.~E., Montgomery, M.~H., \& Williams, K.~A.\ 2010, \apj, 712, 585 
\bibitem[Fontaine \& Brassard(2008)]{Fontaine2008} Fontaine, G., \& Brassard, P.\ 2008, \pasp, 120, 1043 
\bibitem[Gershberg(1972)]{Gershberg1972} Gershberg, R.~E.\ 1972, \apss, 19, 75 
\bibitem[Gilliland et al.(2010a)]{Gilliland2010a} Gilliland, R.~L., Brown, T.~M., Christensen-Dalsgaard, J., et al.\ 2010a, \pasp, 122, 131 
\bibitem[Gilliland et al.(2010b)]{Gilliland2010b} Gilliland, R.~L., Jenkins, J.~M., Borucki, W.~J., et al.\ 2010b, \apjl, 713, L160 
\bibitem[Goldreich \& Wu(1999)]{Goldreich1999} Goldreich, P., \& Wu, Y.\ 1999, \apj, 511, 904 
\bibitem[Greiss et al.(2014)]{Greiss2014} Greiss, S., G{\"a}nsicke, B.~T., Hermes, J.~J., et al.\ 2014, \mnras, 438, 3086 
\bibitem[Hawley et al.(2014)]{Hawley2014} Hawley, S.~L., Davenport, J.~R.~A., Kowalski, A.~F., et al.\ 2014, \apj, 797, 121 
\bibitem[Hermes et al.(2014)]{Hermes2014} Hermes, J.~J., Charpinet, S., Barclay, T., et al.\ 2014, \apj, 789, 85 
\bibitem[Hermes et al.(2015)]{Hermes2015} Hermes, J.~J., Gaensicke, B.~T., Bischoff-Kim, A., et al.\ 2015, arXiv:1505.01848 
\bibitem[Hermes et al.(2011)]{Hermes2011} Hermes, J.~J., Mullally, F., {\O}stensen, R.~H., et al.\ 2011, \apjl, 741, L16 
\bibitem[Hewett et al.(2006)]{Hewett2006} Hewett, P.~C., Warren, S.~J., Leggett, S.~K., \& Hodgkin, S.~T.\ 2006, \mnras, 367, 454 
\bibitem[Hilton(2011)]{Hilton2011} Hilton, E.~J.\ 2011, Ph.D.~Thesis, University of Washington
\bibitem[Hilton(2002)]{Hilton2002} Hilton, J.~L.\ 2002, Asteroids III, 103 
\bibitem[Howell et al.(2014)]{Howell2014} Howell, S.~B., Sobeck, C., Haas, M., et al.\ 2014, \pasp, 126, 398 
\bibitem[Jura(2008)]{Jura2008} Jura, M.\ 2008, \aj, 135, 1785 
\bibitem[Karoff(2014)]{Karoff2014} Karoff, C.\ 2014, \apjl, 781, L22 
\bibitem[Kawaler(2004)]{Kawaler2004} Kawaler, S.~D.\ 2004, Stellar Rotation, 215, 561 
\bibitem[Kepler et al.(2005)]{Kepler2005} Kepler, S.~O., Costa, J.~E.~S., Castanheira, B.~G., et al.\ 2005, \apj, 634, 1311 
\bibitem[Kinemuchi et al.(2012)]{Kinemuchi2012} Kinemuchi, K., Barclay, T., Fanelli, M., et al.\ 2012, \pasp, 124, 963 
\bibitem[Kleinman et al.(1998)]{Kleinman1998} Kleinman, S.~J., Nather, R.~E., Winget, D.~E., et al.\ 1998, \apj, 495, 424 
\bibitem[Koester(2010)]{Koester2010} Koester, D.\ 2010, \memsai, 81, 921 
\bibitem[Koester \& Wilken(2006)]{Koester2006} Koester, D., \& Wilken, D.\ 2006, \aap, 453, 1051 
\bibitem[Lamb \& van Horn(1975)]{Lamb1975} Lamb, D.~Q., \& van Horn, H.~M.\ 1975, \apj, 200, 306 
\bibitem[Landolt(1968)]{Landolt1968} Landolt, A.~U.\ 1968, \apj, 153, 151 
\bibitem[Latter et al.(1987)]{Latter1987} Latter, W.~B., Schmidt, G.~D., \& Green, R.~F.\ 1987, \apj, 320, 308 
\bibitem[Lawlor \& MacDonald(2006)]{Lawlor2006} Lawlor, T.~M., \& MacDonald, J.\ 2006, \mnras, 371, 263 
\bibitem[Lenz \& Breger(2004)]{Lenz2004} Lenz, P., \& Breger, M.\ 2004, The A-Star Puzzle, 224, 786 
\bibitem[Metcalfe(2003)]{Metcalfe2003} Metcalfe, T.~S.\ 2003, Baltic Astronomy, 12, 247 
\bibitem[Moffett(1974)]{Moffett1974} Moffett, T.~J.\ 1974, \apjs, 29, 1 
\bibitem[Mukadam \& Nather(2005)]{Mukadam2005} Mukadam, A.~S., \& Nather, R.~E.\ 2005, Journal of Astrophysics and Astronomy, 26, 321 
\bibitem[Mullally et al.(2008)]{Mullally2008} Mullally, F., Winget, D.~E., De Gennaro, S., et al.\ 2008, \apj, 676, 573 
\bibitem[Nather et al.(1990)]{Nather1990} Nather, R.~E., Winget, D.~E., Clemens, J.~C., Hansen, C.~J., \& Hine, B.~P.\ 1990, \apj, 361, 309 
\bibitem[{\O}stensen et al.(2011)]{Ostensen2011} {\O}stensen, R.~H., Bloemen, S., Vu{\v c}kovi{\'c}, M., et al.\ 2011, \apjl, 736, L39 
\bibitem[Pamyatnykh(1999)]{Pamyatnykh1999} Pamyatnykh, A.~A.\ 1999, Acta Astronomica, 49, 119 
\bibitem[Pfeiffer et al.(1996)]{Pfeiffer1996} Pfeiffer, B., Vauclair, G., Dolez, N., et al.\ 1996, \aap, 314, 182 
\bibitem[Renedo et al.(2010)]{Renedo2010} Renedo, I., Althaus, L.~G., Miller Bertolami, M.~M., et al.\ 2010, \apj, 717, 183 
\bibitem[Saio(2013)]{Saio2013} Saio, H.\ 2013, European Physical Journal Web of Conferences, 43, 05005 
\bibitem[Salaris et al.(1997)]{Salaris1997} Salaris, M., Dom{\'{\i}}nguez, I., Garc{\'{\i}}a-Berro, E., et al.\ 1997, \apj, 486, 413 
\bibitem[Still \& Barclay(2012)]{Still2012} Still, M., \& Barclay, T.\ 2012, Astrophysics Source Code Library, 1208.004 
\bibitem[Tremblay et al.(2013)]{Tremblay2013} Tremblay, P.-E., Ludwig, H.-G., Steffen, M., \& Freytag, B.\ 2013, \aap, 559, A104 
\bibitem[Unno et al.(1989)]{Unno1989} Unno, W., Osaki, Y., Ando, H., Saio, H., \& Shibahashi, H.\ 1989, Nonradial oscillations of stars, Tokyo: University of Tokyo Press, 1989, 2nd ed.,  
\bibitem[Van Grootel et al.(2013)]{VanGrootel2013} Van Grootel, V., Fontaine, G., Brassard, P., \& Dupret, M.-A.\ 2013, \apj, 762, 57 
\bibitem[Walkowicz et al.(2011)]{Walkowicz2011} Walkowicz, L.~M., Basri, G., Batalha, N., et al.\ 2011, \aj, 141, 50 
\bibitem[Wickramasinghe \& Ferrario(2000)]{Wickramasinghe2000} Wickramasinghe, D.~T., \& Ferrario, L.\ 2000, \pasp, 112, 873 
\bibitem[Winget \& Kepler(2008)]{Winget2008} Winget, D.~E., \& Kepler, S.~O.\ 2008, \araa, 46, 157 
\bibitem[Winget et al.(1991)]{Winget1991} Winget, D.~E., Nather, R.~E., Clemens, J.~C., et al.\ 1991, \apj, 378, 326 
\bibitem[Wood(1990)]{Wood1990} Wood, M.~A.\ 1990, Ph.D.~Thesis, The University of Texas at Austin
\end{thebibliography}
\end{document}